УДК 004.415

# СРАВНИТЕЛЬНЫЙ АНАЛИЗ ОСОБЕННОСТЕЙ ИСПОЛЬЗОВАНИЯ В E-LEARNING SVIT И SKYPE

**Малахов К.С., Могильный Г.А., Семенков В.В., Тихонов. Ю.Л, Филипенко С.В.**
Луганский национальный университет им. Тараса Шевченко
кафедра информационных технологий и систем
E-mail: its@luguniv.edu.ua

*Анотация*
*Малахов К.С., Могильный Г.А., Семенков В.В., Тихонов. Ю.Л, Филипенко С.В. Сравнительный анализ особенностей использования в e-learning SVIT и SKYPE.* В работе произведено сравнение возможностей программ SVIT и SKYPE с точки зрения использования в обучении. Исследованы различные типы соединений, качество изображения и звука.

**Общая постановка проблемы.** С развитием технического прогресса в области мультимедийного оборудования стало возможным применение различных способов обучения с использованием компьютерной техники. Для такого способа обучения ввели термин электронное образование (e-learning). E-learning—синоним таких терминов, как электронное обучение, дистанционное обучение, обучение с применением компьютеров, сетевое обучение, виртуальное обучение, обучение при помощи информационных, электронных технологий. В Украине и других странах СНГ, в частности в России, e-learning тоже получило широкое распространение благодаря ряду преимуществ[2]:

- Экстерриториальность — территориальное местонахождение обучающегося не мешает получению информации, не обязателен личный контакт с преподавателем;
- Индивидуальность учебного плана (возможность создать индивидуальные программы обучения, обладающие максимальной актуальностью для каждого конкретного обучающегося);
- Гибкий график (прохождение курса обучения можно организовать в любое время по схеме «24 часа в сутки и 7 дней в неделю»);
- Высокая интерактивность — поиск, диалог, составление сценариев обучения и проверки знаний;
- Мультимедиа — могут быть использованы графические материалы, аудио- и видеоисточники информации и другие мультимедийные возможности;
- Отсутствие «субъективности» экзаменатора — при оценке знаний используются объективные методики оценки и минимизируется напряженность обучающегося, связанная со сдачей очных экзаменов;
- Мониторинг качества обучения — проводится на основе статистического анализа.

Для обеспечения процесса электронного обучения помимо аппаратного обеспечения, требуется также специализированное программное обеспечение (ПО).

Одним из видов таких программ для e-learning является программное обеспечение с закрытым кодом, обеспечивающее шифрованную голосовую связь через интернет между компьютерами – SKYPE[3] и SVIT.

**Цель работы:** провести сравнительный анализ возможностей SKYPE и SVIT в режимах - видеоконференция, чат, транслирование видео. Выявить возможности программ для использования в e-learning.

В целом дистанционное обучение, осуществляемое с помощью компьютерных телекоммуникаций, имеет следующие формы занятий[4]:

*Чат-занятия* — учебные занятия, осуществляемые с использованием чат-технологий. Чат-занятия проводятся синхронно, то есть все участники имеют одновременный доступ к чату. В рамках многих дистанционных учебных заведений действует чат-школа, в которой с помощью чат-кабинетов организуется деятельность дистанционных педагогов и учеников.

*Веб-занятия* — дистанционные уроки, конференции, семинары, деловые игры, лабораторные работы, практикумы и другие формы учебных занятий, проводимых с помощью средств телекоммуникаций и других возможностей «Всемирной паутины».

Для веб-занятий используются специализированные образовательные веб-форумы — форма работы пользователей по определённой теме или проблеме с помощью записей, оставляемых на одном из сайтов с установленной на нем соответствующей программой.

От чат-занятий веб-форумы отличаются возможностью более длительной (многодневной) работы и асинхронным характером взаимодействия учеников и педагогов.

*Телеконференции* — проводятся, как правило, на основе списков рассылки с использованием электронной почты. Для учебных телеконференций характерно достижение образовательных задач. Также существуют формы дистанционного обучения, при котором учебные материалы высылаются почтой в регионы.

В основе такой системы заложен метод обучения, который получил название «Природный процесс обучения» (Natural Learning Manner). Дистанционное обучение — это демократичная простая и свободная система обучения [5]. Она была изобретена в Великобритании и сейчас активно используется жителями Европы для получения дополнительного образования. Система дистанционного обучения - образовательная система, обеспечивающая получение образования с помощью дистанционных технологий обучения. Студент, постоянно выполняя практические задания, приобретает устойчивые автоматизированные навыки. Теоретические знания усваиваются без дополнительных усилий, органично вплетаясь в тренировочные упражнения. Формирование теоретических и практических навыков достигается в процессе систематического изучения материалов и прослушивания и повторения за диктором упражнений на аудио и видео-носителях (при наличии).

[On-line-конференция](#) является эффективным инструментом обмена информацией. Она расширяет возможности для профессиональной деятельности, образования и науки, общения и творчества.

**Обычно видеоконференция** включает в себя следующие модули:[6]
- общий и приватный текстовый чат;
- список пользователей с различными функциями и полномочиями;
- рабочая область («Доска»), включающая панель управления презентациями и панель инструментов;
- видео модуль с «трибунами» для выступающих.

On-line конференция — это эффективное средство получения качественного и доступного образования.

Одним из главных требований к системам используемым для обучения является наличие формализованной методики, действенной информационной и функциональной модели и соответствующих инструментальных средств программной реализации. При этом должен использоваться в качестве источника знаний выступает лингвистический корпус текстов множества ОУ на русском и украинском языке из заданной предметной дисциплины. Для обработки текстов на украинском языке могут использоваться лингвистические программные процессоры и способы обработки больших объемов текстовой информации во времени, приближенном к реальному.

Рассмотрим программу SKYPE, которая сейчас широко используется в мире. Для работы в SKYPE необходимы: микрофон, динамики, веб-камера, точка доступа для подключения к интернету. Для более эффективной и наглядной работы можно использовать мультимедийную доску и проектор.

В SKYPE имеется возможность осуществлять голосовые и видео-звонки звонки.

При использовании функции «видеозвонок» появляется окно с изображением с камеры абонента в специальном окне.

SKYPE обеспечивает возможность:

- Проведения on-line лекций;
- Передачи необходимых медиа-файлов (с текстами, пояснениями к лекциям, презентации);
- Передачи изображения со своего экрана (режим чтения экрана);
- Получения ответов преподавателей на вопросы студентов on-line.

Рассмотрим альтернативную программу SVIT. Сначала необходимо произвести ее настройку. В SVIT после выполнения команд меню «Опции», «Сеть» откроется окно, в котором производится настройка параметров соединения.

Как и в SKYPE необходимо создать учетную запись (команда «Опции», «Учетная запись»).

SVIT обеспечивает те же возможности, что и SKYPE (Проведения on-line лекций, передачи необходимых медиа - файлов, трансляции обучающего видео, получения ответов преподавателей на вопросы студентов on-line.

Дополнительно предоставляется возможность работать с интерактивной доской. Для вызова интерактивной доски выполняем команды «Модули», «Интерактивная доска». Откроется окно «Виртуальная доска». Выбираем команды «Файл», «Новый». Для работы с доской предусмотрены различные инструменты (перо, маркер, различные геометрические фигуры). Для построения различных эскизов, схем есть возможность набора текста и вставки графических объектов. В программе предусмотрена возможность трансляции видео файлов с расширением AVI (Audio Video Interleave). Формат файлов с расширением AVI может содержать видео и аудио данные, сжатые с использованием разных комбинаций кодеков, что позволяет синхронно воспроизводить видео со звуком. AVI-файлы могут содержать различные виды сжатых данных, к примеру DivX для видеоинформации и MP3 для аудио. Для трансляции необходимо выбрать меню «Модули», «Транслировать файл».

В отличие от SKYPE, программа SVIT позволяет соединяться через локальную сеть. Для настройки такого типа соединения требуется ввести IP адрес компьютера и порт.

В процессе работы осуществлялось тестирование SVIT в режимах -видеоконференция, чат, виртуальная доска, транслирование видео. Тестирование SKYPE проводилось в режимах- видеоконференция, чат, транслирование видео.

В видеоконференциях в обеих системах в условиях подключения SVIT через локальную сеть с использованием технологии NAT (Network Address Translation— «преобразование сетевых адресов») и SKYPE через локальную сеть посредством прокси – сервера (служба в [компьютерных сетях](#), позволяющая [клиентам](#) выполнять косвенные запросы к другим сетевым службам.) со скоростью 1 Мб/с получен одинаковый результат по скорости передачи изображения и звука, без заметных задержек. В режиме чат, при тех же условиях подключения, в обеих системах скорость прохождения текста показала одинаковый результат.

Трансляция видео при тех же условиях подключения в обеих системах проходила с небольшими задержками.

Однако, в настройках SKYPE имеется возможность подключения к прокси-серверу, а для подключения SVIT приходится дополнительно, настраивать систему NAT. Кроме того, в отличие от SVIT, SKYPE имеет возможность передавать окна других приложений.

В режиме виртуальной доски работает только SVIT. Это дает дополнительные возможности для разъяснения учебного материала по ходу его изучения. Например, виртуальная доска позволит проводить совместное построение онтографа (Рис.1). Таким образом, SVIT является программным обеспечением для совместной работы. Исходя из этого, есть возможность осуществлять взаимодействия между людьми, совместно работающими над решением общих задач. С помощью типа такой коллективной работы достигается высокая эффективность и увеличивается скорость получения конечного требуемого результата. Так как устоявшегося русскоязычного термина пока нет, описывающего тип данного взаимодействия, часто используется один из англоязычных без перевода (collaboration, workgroup) [7].

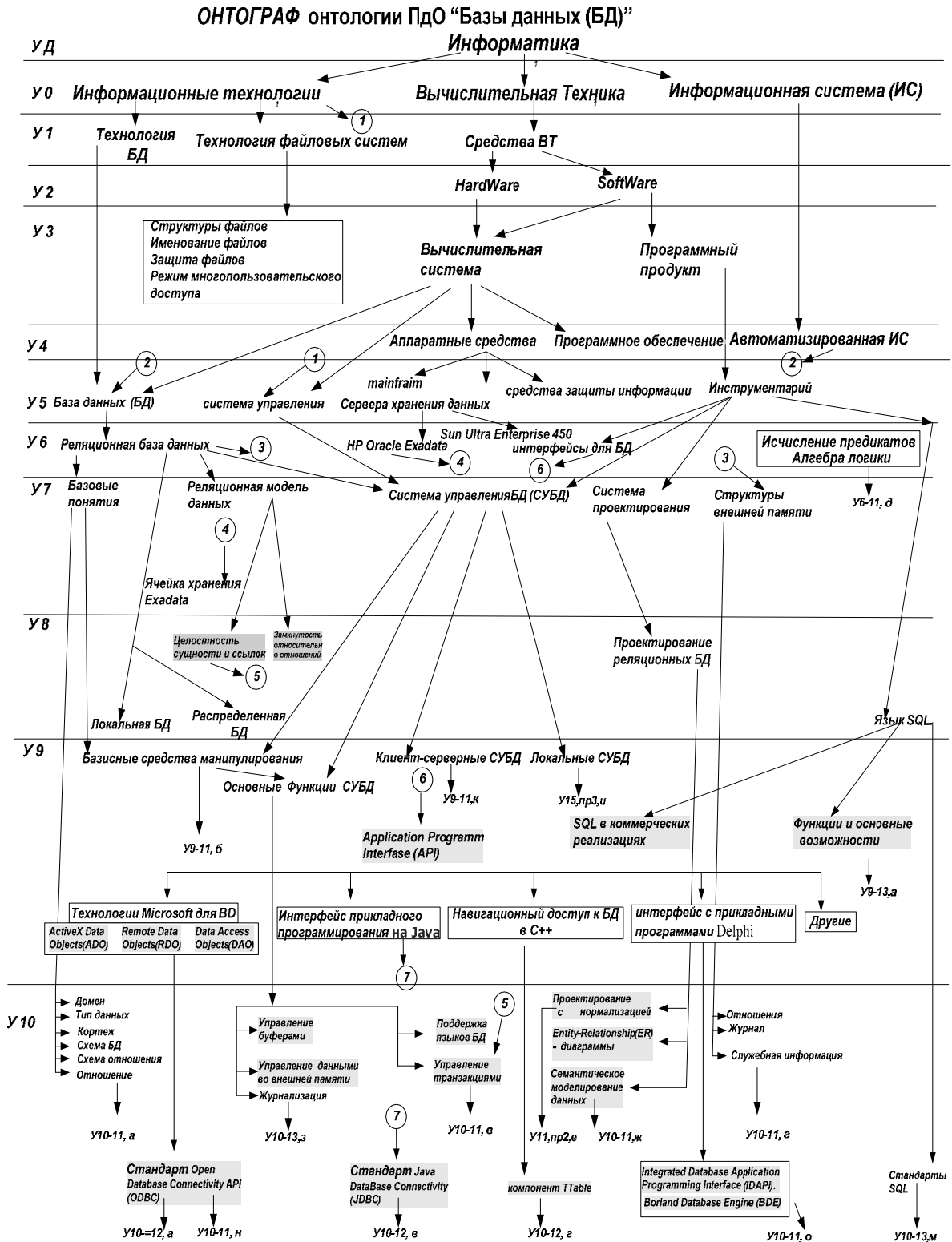

Рис 1. Онтограф онтологии ПдО «Базы данных (БД)»

**Выводы**

В результате тестирования функциональных возможностей программ были выявлены некоторые особенности. Трансляция рисунков и надписей происходила без длительных задержек в двустороннем режиме. Возникли небольшие проблемы с транслированием текстового документа. Документ отображался с некоторой задержкой, которая зависит от скорости Интернет-соединения.

В целом выяснилось, что в режимах видеоконференция, чат, транслирование видео обе системы равноценны для e-learning. Однако, SVIT обладает дополнительным режимом - виртуальная доска, которая позволяет давать дополнительные разъяснения материала on-line. В SVIT так же предусмотрен режим подключения по локальной сети без использования подключения через интернет. Кроме того, в SVIT можно одновременно транслировать и виртуальную доску, и изображение с веб – камеры. SKYPE имеет преимущество – режим чтения экрана (передача изображения с экрана).

При условии доработки для получения в настройках возможности соединение через прокси-сервер и возможности передавать окна других приложений (режим чтения экрана) SVIT становится хорошим альтернативным ПО для электронного обучения.

**Литература**